\documentclass{PoS}
\usepackage{caption}
\usepackage{subcaption}
\title{The intrinsic Baldwin effect in NLSy 1 galaxies}

\ShortTitle{Intrinsic Beff in NLSy1}

\author{{Nemanja Raki\'c,}$^{ab}$ Dragana Ili\'c$^b$ and Luka \v{C}. Popovi\'c$^{abc}$ \\
       \llap{$^a$}Faculty of Natural Sciences and Mathematics, University of Banjaluka, Mladena Stojanovi\'ca 2, 78000 Banjaluka,  Republic of Srpska, Bosnia and Herzegovina\\
       \llap{$^b$}Faculty of Mathematics, University of Belgrade, Studentski Trg 16, 11000 Belgrade, Serbia\\
       \llap{$^c$}Astronomical observatory Belgrade, 11060 Belgrade, Serbia\\
        E-mail: \email{nemanja.rakic@pmf.unibl.org}, \email{dilic@matf.bg.ac.rs}, \email{lpopovic@aob.rs}}

\abstract{The intrinsic Baldwin effect is an anti-correlation between the line equivalent
width and the flux of the underlying continuum detected in a single variable
active galactic nucleus (AGN). This effect, in spite of the extensive research,
is still not well understood, and might give us more information about the
physical properties of the line and continuum emission regions in AGNs. Here we
present preliminary results of our investigation of the intrinsic Baldwin
effect of the broad H$\beta$ line in several Narrow-line Seyfert 1 (NLSy 1) galaxies, for which data
were taken from the long-term monitoring campaigns and from the Sloan Digital
Sky Survey Reverberation Mapping project. }

\FullConference{Revisiting narrow-line Seyfert 1 galaxies and their place in the Universe - NLS1 Padova\\
		9-13 April 2018 \\
		Padova Botanical Garden, Italy}

\begin{document}

\section{Introduction}
Broad line region (BLR) of active galactic nuclei (AGNs) is found relatively
near to the energetic source of ionization that comes from the accretion disk,
thus the conditions of the BLR gas are such that it is hard to compare them
with those in other well studied astrophysical objects. As a consequence many
of the standard techniques, previously derived to probe the physics of e.g.
photoionized nebulae, are often unable to give reliable results when applied to
the BLR.
 The intrinsic Baldwin effect (Beff) is an anti-correlation between the emission line
equivalent width and the flux of the underlying continuum detected in a single
variable AGN \cite{bal77}. Recently, we studied
intrinsic Beff on sample of six type 1 AGNs, including two Seyfert 1, two AGNs
with very broad double-peaked lines, one NLSy1 galaxy and one high-luminosity
quasar \cite{ja17}. We found that all six galaxies
exhibit intrinsic Beff, including the NLSy1 Ark 564, and that this effect is
probably not caused by the geometry of the broad line region (BLR).
Additionally, we showed that there is no connection between global and
intrinsic Beff \cite{ja17}.

Taking into account our first small sample of only 6 objects (with only one
NLSy1), our findings should be tested on a larger number of different AGNs.
Therefore, we aim to study the intrinsic Beff on a sample of type 1 AGN taken
from the Sloan Digital Sky Survey Reverberation Mapping project (SDSS-RM). Here
we present our preliminary results of our study of the intrinsic Beff of broad H$\beta$
line in five NLSy1 galaxies from SDSS-RM, compared to the findings in Ark 564
from \cite{ja17}.
\section{Data and Methods}
 We selected objects from SDSS-RM explained in \cite{sdssrm} with $z<0.8$ and we cross-matched with NLSy 1 catalog of \cite{Rak17}. Additionally, we used the data of long-term monitored NLSy1 galaxy Ark 564 published in \cite{ja17,sh12}. The information on selected objects is given in Table \ref{tab:gal}.
\begin{table}[h!]

\centering
\scalebox{0.8}{
\begin{tabular}{c c c c c c c c}
\hline\hline
Name& RA & DEC & z & Period & No. of Spectra &S/N& Our notation  \\
(1)&(2)&(3)&(4)&(5)&(6)&(7)&(8)\\
\hline
J141308.10+515210.3&213.28377&51.86955&0.288&56660--57518&48&13&N1\\
J141721.80+534102.6&214.34082&53.68406&0.193&56660--57518&48&33&N2\\
J141427.88+535309.6&213.6162&53.88602&0.242&56660--57518&48&10&N3\\
J141419.84+533815.3&213.58268&53.6376&0.164&56660--57518&48&45&N4\\
J141408.76+533938.2&213.5365&53.66063&0.191&56660--57518&48&23&N5\\
Ark 564&340.6625&29.72530&0.025&51424--54414&92&>50&--\\

\hline
\end{tabular}
}
\caption{Selected NLSy 1 galaxies. The columns are denoted in following way: (1) Name of the galaxy; (2) R.A. in degrees; (3) DEC. in degrees; (4) redshift; (5) monitored period in MJD; (6) number of spectra;(7) averaged signal to noise; (8) notation used in this paper\label{tab:gal}}

\end{table}

We rescale spectra taking that the  flux of [OIII]$\rm \lambda 5007$ remained constant within the monitored period \cite{pe93}.
To remove the host galaxy contribution  we used Simple Stellar Population method (SSP) \cite{Rak17,zh14}. With this method observed spectrum $F(\lambda)$ (where we used median smoothing to mask emission lines) can be approximated with:

$$F(\lambda)=\left[\sum_{n=1}^{39} a_n \times F_{ssp}(\lambda)\right] \otimes G(\sigma)+F_{AGN}(\lambda).$$
The left part of the previous equation represents the host galaxy contribution, where $a_n$ is the amplitude of individual SSP template ($F_{ssp}$), which is broadened with Gaussian broadening function $G(\sigma)$ ($\sigma$ being stellar velocity dispersion), while, on the right side,  AGN is represented as $F_{AGN}$ in the form of the power law $\lambda^\alpha$.  

In order to subtract narrow and satellite lines and extract the broad component, we simultaneously fitted emission lines including the local continuum, in the wavelength region from 4200 \AA\ to 5500 \AA. Lines included in the fit were: H$\gamma$ (two components: broad and narrow), HeII (broad and narrow), H$\beta$ (broad and narrow), forbidden lines [OIII]$\lambda$4363,4959,5007 \AA\ and the FeII template\footnote{Fe II template is available on \href{http://servo.aob.rs/FeII_AGN/}{http://servo.aob.rs/FeII\_AGN/}} from \cite{kov10} and \cite{sh12}. For further investigation of the intrinsic Beff we followed methods from \cite{ja17}.
\section{Results and Discussion}
\begin{figure}[h!]
\centering
                    \includegraphics[width=.9\linewidth]{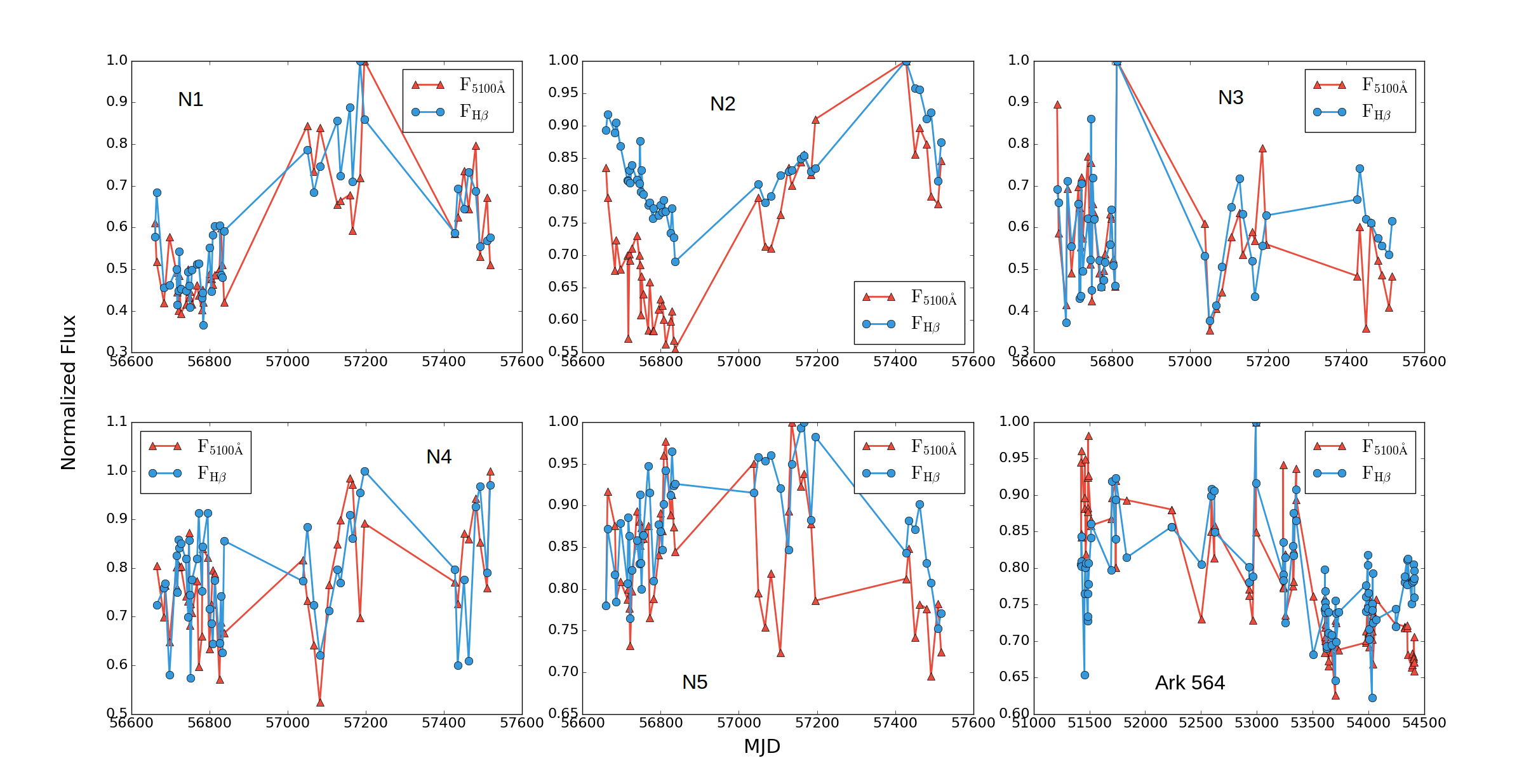}
                    \caption{Light curves of 5 NLSy1 galaxies from SDSS-RM project with Ark 564. All fluxes are normalized to the maximum value. The light curve of continuum is represented with red triangles, while light curve of H$\beta$ with blue circles. Names of objects are denoted on the plots.}
                    \label{lc}
				\end{figure}
In Fig. \ref{lc} the light curves of the continuum and H$\beta$ fluxes of the five NLSy1 galaxies selected from SDSS-RM are shown together with the light curves of Ark 564 \cite{ja17}. In order to compare changes in the light curve of $F(\rm{H}\beta)$ with corresponding changes in light curves of $F_{5100}$ we show fluxes normalized to the maximum value.	It is straightforward to see that the flux of H$\beta$ is not strictly following the changes in the continuum flux, e.g. where we have local minimum in $F(\rm{H}\beta)$ we have a local maximum in $F_{5100}$ (see for e.g. light	curve of N5 in Fig. \ref{lc}) and vice versa. This might suggest the existence of an additional continuum emission, which we previously discussed in \cite{ja17}.
          
\begin{figure}[h!]
				\centering
                    \includegraphics[width=.9\linewidth]{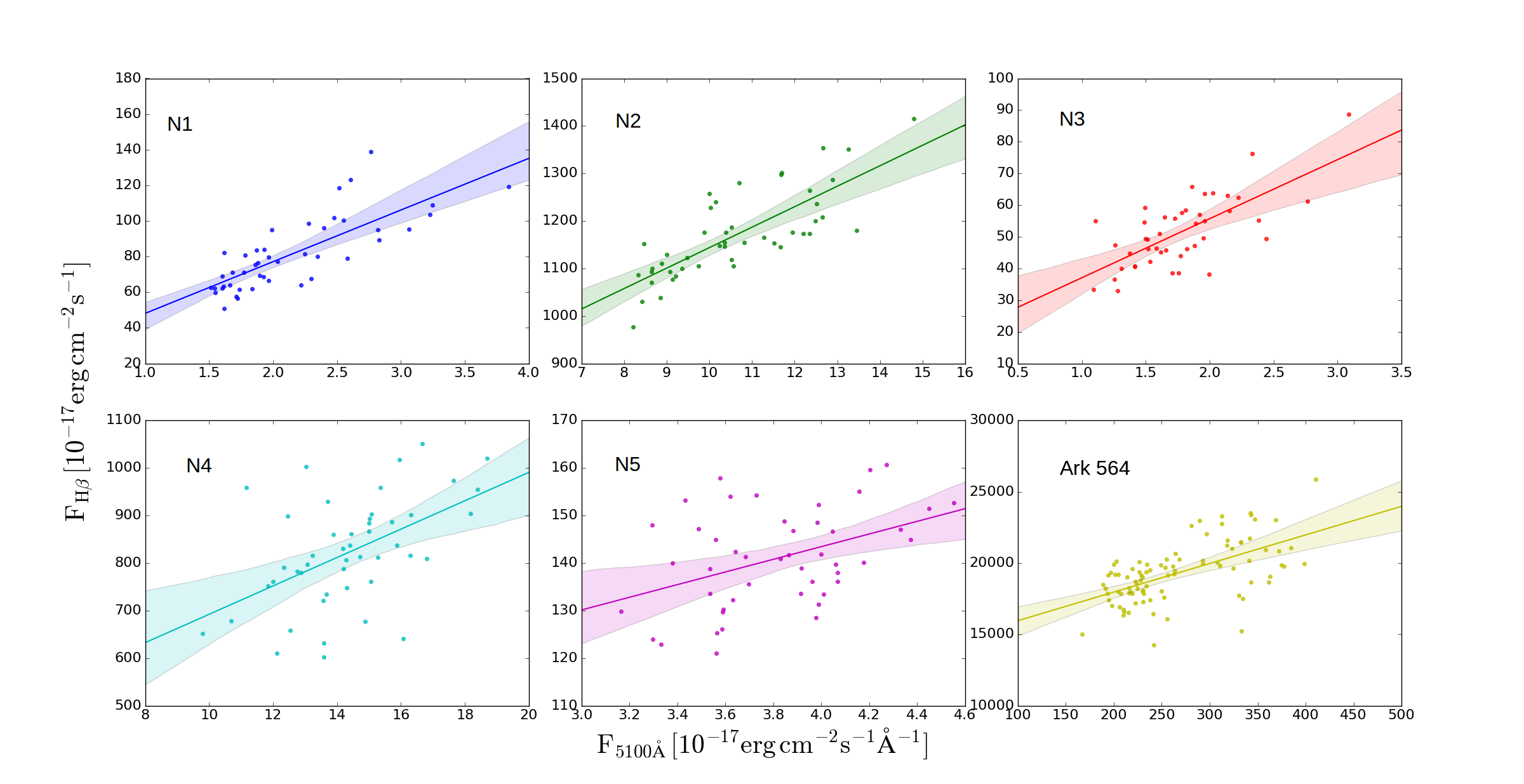}
                    \caption{Flux of continuum at 5100 \AA\ vs. flux of the broad H$\beta$ line. The linear best fit is presented with solid line, while shaded area represents confidence interval of 95 \%. The names are denoted on plots.}
                    \label{fvsf}
				\end{figure}

In Fig. \ref{fvsf} we showed relationship between continuum flux at 5100 \AA\ and flux of the H$\beta$.	The strongest correlation (0.782) is found in N1 object, while weakest in N5 object (0.442). In case of all objects there is the linear correlation between the line and continuum flux. 			
				
\begin{figure}[h!]
				\centering
                    \includegraphics[width=.9\linewidth]{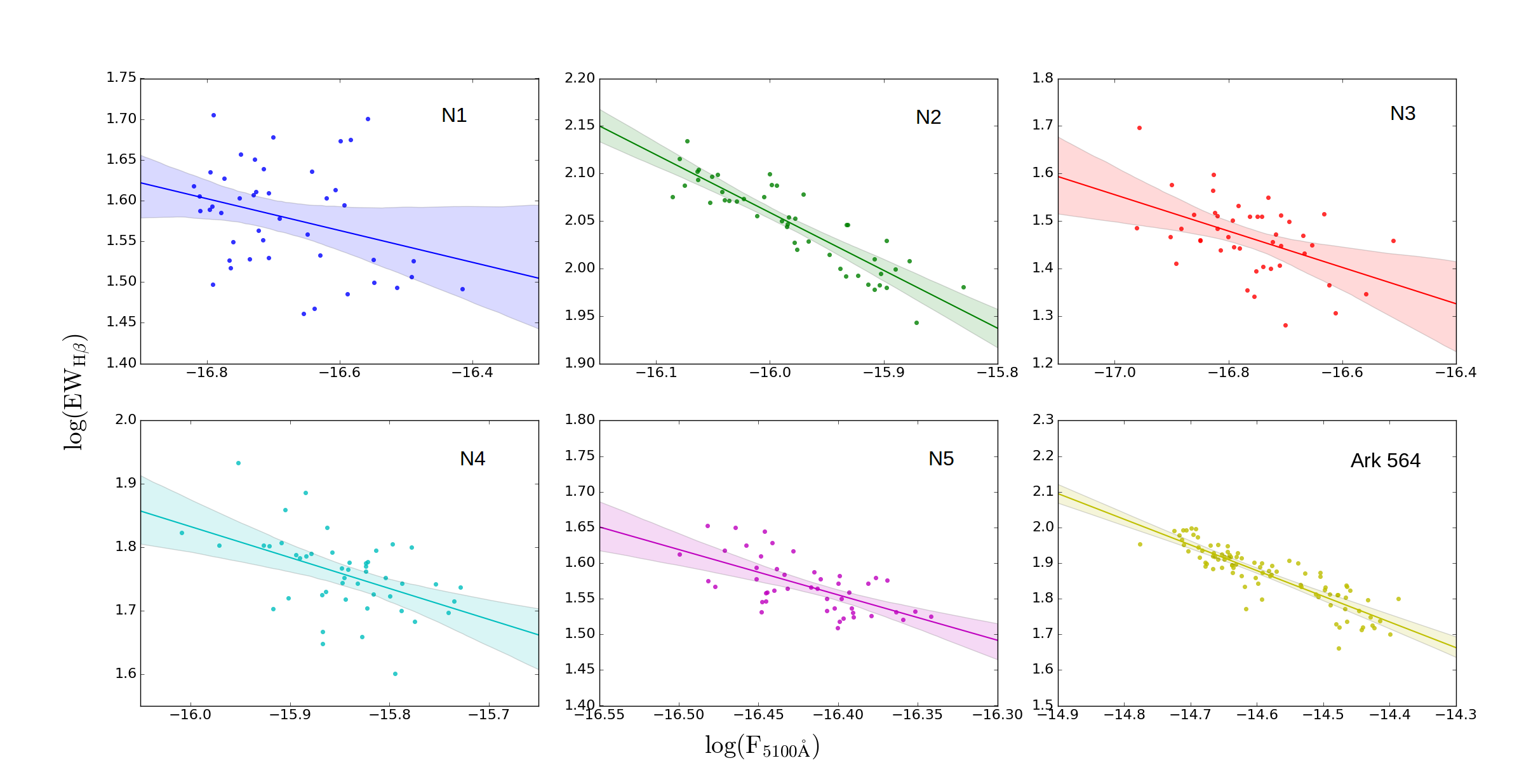}
                    \caption{Intrinsic Beff for the sample of five AGNs form SDSS-RM project together with Ark 564. The solid line represents the best linear fit, while shaded area corresponding confidence interval of 95\%.}
                    \label{bald}
				\end{figure}

In Fig. \ref{bald} we present the intrinsic Beff of selected galaxies together with Ark 564. We found that all  galaxies except N1 are exhibiting intrinsic Beff. Calculated Pearson correlation of line flux versus continuum flux and parameters of linear fit together with the calculated Pearson correlation for the intrinsic Beff  for all considered objects are given in the Table \ref{tab:bald}. Interestingly, in Ark 564 there is a weak correlation between line and continuum flux, but there is the strongest intrinsic Beff present. We note that  Ark 564 has been monitored for 11 years, while SDSS galaxies for only about 2.5 years, this might be the reason behind weaker correlation found in SDSS galaxies.It is also important to emphasize that in the case of SDSS objects, line fluxes
are not corrected for the time delay with respect to the continuum.
However, we do not expect this will have significant impact on the
Baldwin effect results, since in NLSy1 the time delay is small and the
variability is weak, e.g. we tested this effect in case of Ark 564 which has
the time delay of $\sim$5 days and weak variability of $\sim$ 10\% \cite{sh12}, and there is no
difference in the result. But in general, it is necessary to determine EW so that it represents the real emission line response to the continuum that actually hit the BLR at time when the line was emitted (see \cite{ja17,gp03,gk14}). We notice that objects showing weaker correlation between line and continuum
flux, are on the other hand having stronger intrinsic Beff. It may be that the
stronger intrinsic Beff is explained with the presence of an additional
continuum emission, as previously discussed in \cite{ja17}.

                      	\begin{table}[]
                      	\small
\centering
\caption{Parameters of the intrinsic Baldwin effect }
\label{tab:bald}
\scalebox{0.8}{
\begin{tabular}{cccc ccccc}
\hline\hline
Object&\multicolumn{4}{c}{Line flux vs. cnt. flux}&\multicolumn{4}{c}{Intrinsic Beff}\\

Name&\multicolumn{2}{c}{Pearson correlation}&\multicolumn{2}{c}{Linear fit}&\multicolumn{2}{c}{Pearson correlation}&\multicolumn{2}{c}{Linear fit}\\
& r & P&slope&intercept&r&P&$\beta$&A\\\hline
J141308.10+515210.3&0.782&$3\times 10^{-10}$&29.0&2$\times 10^{-16}$&-0.309  & $4\times 10^{-2}$ &-0.195 &-1.677    \\
J141721.80+534102.6&0.781&$1\times10^{-10}$&43.1&7$\times10^{-15}$&-0.891 & $9\times10^{-17}$&-0.608&-7.671 \\
J141427.88+535309.6&0.697&$1\times10^{-7}$&18.6&2$\times 10^{-16}$&-0.483 & $1\times10^{-4}$&-0.382&-4.946 \\
J141419.84+533815.3&0.5158&$2\times10^{-4}$&29.9&4$\times10^{-15}$&-0.484 & $5\times10^{-4}$ &-0.498&-5.990  \\
J141408.76+533938.2&0.442&$3\times10^{-3}$&13.3&9$\times10^{-16}$&-0.635 & $2\times10^{-6}$&-0.636&-8.875\\ 
Ark 564 &0.592&$7\times10^{-10}$&20.1&$10^{4}$&-0.879&$2\times 10^{-30}$&-0.721&-8.642\\\hline
\end{tabular}
}
\end{table} 
\section*{Acknowledgements}

This conference has been organized with the support of the
Department of Physics and Astronomy ``Galileo Galilei'', the 
University of Padova, the National Institute of Astrophysics 
INAF, the Padova Planetarium, and the RadioNet consortium. 
RadioNet has received funding from the European Union's
Horizon 2020 research and innovation programme under 
grant agreement No~730562. This  paper is  based  upon  work  from  COST  Action
CA16104  "GWverse",  supported  by  COST  (European cooperation in Science and Technology). This work is a part of the project (176001)
"Astrophysical Spectroscopy of Extragalactic Objects", supported
by the Ministry of Education, Science and Technological Development of Serbia and the project "Investigation of super-massive binary black holes in the optical and X-ray spectra"
supported by the Ministry of Science and Technology of R.
Srpska.

\bibliographystyle{JHEP}
\bibliography{nlsy1}

\end{document}